 \documentstyle[12pt]{article}
 \def\be{\begin{eqnarray}}
 \def\ee{\end{eqnarray}}
 \def\ba{\begin{array}}
 \def\ea{\end{array}}

\begin{document}
 \begin{center}
 {\bf\Large
 {Kalb--Ramond dipole solution in\\
 \vskip 3mm
 low-energy bosonic string theory}}
 \end{center}

 \vskip 1cm

 \begin{center}
 {\bf \large {Alfredo Herrera-Aguilar}}
 \end{center}

 \begin{center}
 Instituto de F\'\i sica y Matem\'aticas\\
 Universidad Michoacana de San Nicol\'as de Hidalgo\\
 Apdo. Postal 2-82, Morelia, Mich., M\'exico\\
 e-mail: herrera@zeus.umich.mx
 \end{center}

 \vskip 3mm
 \begin{center}
 and
 \end{center}
 \vskip 5mm

 \begin{center}
 {\bf \large {Oleg V. Kechkin}}
 \end{center}

 \begin{center}
 Institute of Nuclear Physics,\\
 M.V. Lomonosov Moscow State University, \\
 Vorob'jovy Gory, 119899 Moscow, Russia, \\
 e-mail: kechkin@depni.npi.msu.su
 \end{center}

 \vskip 1cm
 \begin{abstract}
We construct a new solution subspace for the bosonic string theory
toroidally compactified to 3 dimensions. This subspace corresponds
to the complex harmonic scalar field coupled to the effective
3--dimensional gravity. We calculate a class of the asymptotically
flat and free of the Dirac string peculiarity solutions which describes
a Kalb--Ramond dipole source with the generally nontrivial dilaton charge.
 \end{abstract}

 \newpage

 \section{Introduction}

The main part of the valid string theory results was obtained using a
perturbative approach \cite{kir}. Moreover, the most nonperturbative
information about string theory was derived from the perturbative one
by the help of the conjectured nonperturbative string dualities \cite{st}.
In the framework of the perturbative approach one reduces the
quantum string dynamics at low energies to the classical field theory of
the string low mass exhitation modes \cite{gsw}. A solution spectrum of
these effective string theories provide the main tool for the following
quantum string theory investigation \cite{y}.

The field theories of the string exhitation modes are essentially nonlinear.
To construct wide classes of exact solutions of these theories
one must use both the consistent ansatze search and the symmetry technique
application \cite{pde}. The best strategy for generation of the new classical
solutions consists of the straightforward construction of the simple solution
of the interested type with its following generalization by the help of the
symmetry transformations. Finally, one obtains a symmetry
invariant solution class of the same type if the symmetries used for the
generalization form a subgroup preserving the type under consideration
(see \cite{cs} for the asymptotically flat fields in string theories).

In this paper we deal with the effective field theory describing the
low--energy dynamics of the bosonic string theory massless exhitation modes.
These modes include the scalar dilaton field and the tensor
Kalb--Ramond and graviton (metric) ones living in the multidimensional
space--time. The equations of the theory have a highly nonlinear form and
correspond to the multidimensional General Relativity \cite{w} with the
nontrivially coupled dilaton and Kalb--Ramond matter fields. In this paper,
using the ansatze approach, we construct a class of the asymptotically flat
solutions which possesses the invariance property under the action of the
symmetry subgroup preserving the ansatze taken.

The paper is organized as follows: in section 2 we define a consistent
truncation of the bosonic string theory and reformulate the resulting
system in terms of the matrix--valued symmetric space model \cite{bgm}. In
section 3 we construct an ansatze which corresponds to the complex harmonic
scalar field coupled to the effective gravity in 3 dimensions. In section 4
we construct a special axisymmetric solution of the ansatze equations which
is defined by the harmonic function of the Coulomb form with the complex
``charge'' and the imaginary ``location'' on the symmetry axis. The
corresponding solution of the bosonic string theory describes a Kalb--Ramond
dipole with the dilaton charge. In the general case it possesses a
Dirac string peculiarity in the general case; we establish all the
special situations where this peculiarity vanishes. In Conclusion we discuss
the possible symmetry generalization of the constructed solution in the
framework of the bosonic and heterotic string theories and also consider
the general perspectives of the approach applied in this paper.


\section{A truncation}

Let $X^{M}$ ($M=1,...,D$) be the coordinates of the space--time with the
signature ($-+ \cdots +$). Let $\Phi$, $B_{MN}=-B_{NM}$ and $G_{MN}=G_{NM}$
denote the dilaton, Kalb--Ramond and metric fields correspondingly. Then the
action of the classical field theory which describes the low--energy dynamics
of these massless modes of the bosonic string theory reads \cite{gsw}:
\be\label{1}
S_D=\int d^DX\sqrt{-{\rm det}G_{MN}}e^{-\Phi}\left ( R_D+\Phi_{,M}\Phi^{,M}-
\frac{1}{12}H_{MNK}H^{MNK}
\right ),
\ee
where $H_{MNK}=\partial_MB_{NK}+\partial_KB_{MN}+\partial_NB_{KM}$. In this
paper we deal with the special truncation of the theory (\ref{1}) which can be
performed using two steps. Let us put $D=d+3$ and denote $y^m=X^m$
($m=1,...,d$) and $x^{\mu}=X^{d+\mu}$ ($\mu=1,2,3$). We impose the
following consistent restrictions at the first step:
\be\label{2}
B_{mn}=G_{m,d+\mu}=0.
\ee
The remaining nontrivial field components consist of $G_{mn}, B_{m,d+\mu}$ (we
combine them into the $d\times d$ and $d\times 1$ matrices $G$ and $b_{\mu}$
correspondingly), and also the fields $\Phi$, $G_{d+\mu, d+\nu}$ and
$B_{d+\mu, d+\nu}$. At the second step we suppose that these fields are
independent on the coordinates $y^m$, i.e. we perform the toroidal
compactification of the first $d$ dimensions \cite{ms}, \cite{s}. The
resulting dynamical system admits a consistent restriction
\be\label{3}
B_{d+\mu, d+\nu}=0,
\ee
and can be naturally expressed in terms of the fields $G,\phi,v$ and
$h_{\mu\nu}$, where
\be\label{4}
\phi=\Phi-{\rm ln}\sqrt{-{\rm det}G},
\ee
\be\label{5}
h_{\mu\nu}=e^{-2\phi}G_{d+\mu,d+\nu},
\ee
and $v$ is defined on shell by the differential relation
\be\label{6}
\nabla v=e^{-2\phi}G^{-1}\nabla\times\vec b.
\ee
In Eq. (\ref{6}) $(\vec b)_{\mu}=b_{\mu}$; also in this equation and below all
the differential operations are related to the coordinates $x^{\mu}$ and the
3-dimensional indeces are lowered and raised using the metric $h_{\mu\nu}$ and
its inverse one $h^{\mu\nu}$. Finally the motion equations for the truncated
theory of the bosonic string take the following form:
\be\label{7}
\nabla J=0,
\ee
\be\label{8}
R_{3\,\,\mu\nu}=\frac{1}{4}{\rm Tr} \left ( J_{\mu}J_{\nu}\right ),
\ee
where the Ricci tensor $R_{3\,\,\mu\nu}$ is constructed using the 3--metric
$h_{\mu\nu}$, $J=\nabla {\cal G}\,{\cal G}^{-1}$ and
\be\label{9}
{\cal G}=\left (\ba{ccc}
-e^{-2\phi}+v^TGv&\,\,&v^TG\cr
Gv&\,\,&G
\ea\right ),\quad
{\cal G}^{-1}=\left(\ba{ccc}
-e^{2\phi}&\,\,&e^{2\phi}v^T\cr
e^{2\phi}v&\,\,&G^{-1}-e^{2\phi}vv^T
\ea\right ).
\ee
Using a solution of the system (\ref{7})--(\ref{8}) and performing Eq.
(\ref{9}) one can easily calculate the multidimensional dilaton field $\Phi$
and the metric $ds_{d+3}^2$; the result reads:
\be\label{10}
e^{\Phi}&=&\sqrt{{\rm det}{\cal G}}\,e^{2\phi}\nonumber\\
ds_{d+3}^2&=&dy^TGdy+e^{2\phi}ds_3^2,
\ee
where $y$ is the $d\times 1$ column with the components $y^m$ and
\be\label{10'}
ds_3^2=h_{\mu\nu}dx^{\mu}dx^{\nu}.
\ee
From Eq. (\ref{9}) it follows that
\be\label{11}
G={\cal G}_{22},\quad e^{2\phi}=-\left ({\cal G}^{-1}\right )_{11},
\ee
where the indeces enumerate the corresponding matrix blocks. To calculate the
Kalb--Ramond field component $B_{m,d+\mu}$ it is convenient to introduce the
vector matrix potential $\vec\Omega$ accordingly the relation
\be\label{12}
\nabla\vec\Omega=J.
\ee
This potential exists on shell of Eq. (\ref{7}); using its definition and Eq.
(\ref{6}) it is easy to prove that
\be\label{13}
B_{m,d+\mu}=-\left (\vec\Omega_{21}\right )_{m\mu}.
\ee

Thus, the strategy for construction of a concrete solution for the theory
(\ref{1}) truncated accordingly the presented procedure consists of the
calculation of the quantities ${\cal G}, {\cal G}^{-1}, {\rm det}{\cal G}$ and
$\vec\Omega$ and of the following application of Eqs. (\ref{10}), (\ref{11})
and (\ref{13}). Before this application one can generalize the obtained
solution using the transformation
\be\label{14}
{\cal G}\rightarrow {\cal C}^T{\cal G}{\cal C},\quad \vec\Omega\rightarrow
{\cal C}^T\vec\Omega{\cal C}^{T\,\,-1}
\ee
which gives a symmetry of the theory, see Eqs. (\ref{7}), (\ref{8}) and
(\ref{12}).


\section{A subspace}

The matrix ${\cal G}$ introduced in the previous section must be a symmetric
nondegenerated matrix of the signature $(--+\cdots +)$. The simplest solution
of the motion equations (\ref{7})--(\ref{8})
possessing these properties is the matrix ${\cal G}_0={\rm diag}
(-1,-1,1,...,1)$; the corresponding 3--metric $h_{\mu\nu}$ will describe the
flat Euclidean 3--space. Let us now consider an ansatze with
\be\label{15}
{\cal G}={\cal G}_0+\sum_k\lambda_kC_kC_k^T,
\ee
where $\lambda_k=\lambda_k(x^{\mu})$ are the (nonmatrix) functions, $C_k$ are
the constant columns restricted by the relations
\be\label{16}
C_k^T{\cal G}_0C_l=\sigma_k\delta_{kl}
\ee
where $\sigma_k=0,\pm 1$, and $k=1,...,{\cal N}$. This ansatze includes the
simplest solution mentioned above for the choice $\lambda_k=0$ and provides
the necessary matrix ${\cal G}$ properties at least for the small $\lambda_k$
values. Eq. (\ref{16}) means that the columns $C_k$ are normalized
and mutually orthogonal in respect to the ``metric'' ${\cal G}_0$. The
normalization conditions do not restrict the ansatze generality because the
nonzero norms $C_k^T{\cal G}_0C_k$ can be absorbed by the functions
$\lambda_k$, as it is seen from Eq. (\ref{15}). More precisely, the column
$C_k$ restricted by Eq. (\ref{16}) is normalized for $\sigma_k=\pm 1$ and
selforthogonal for $\sigma_k=0$.

Using the relations (\ref{16}) one can calculate the matrices ${\cal G}^{-1}$
and $J$. The result reads:
\be\label{17}
{\cal G}^{-1}={\cal G}_0+\sum_k\mu_k{\cal G}_0C_kC_k^T{\cal G}_0,
\ee
where
\be\label{18}
\mu_k=-\frac{\lambda_k}{1+\sigma_k\lambda_k}
\ee
and
\be\label{19}
J=\sum_k\frac{\nabla\lambda_k}{1+\sigma_k\lambda_k}C_kC_k^T{\cal G}_0.
\ee
Let us now introduce the functions $\xi_k$ as
\be\label{20}
\xi_k=\int_0^{\lambda_k}\frac{d\lambda}{1+\sigma_k\lambda}.
\ee
Then for the functions $\lambda_k$ and $\mu_k$ one obtains the following
symmetric expressions:
\be\label{21}
\lambda_k=\frac{e^{\sigma_k\xi_k}-1}{\sigma_k}, \quad \mu_k=
\frac{e^{-\sigma_k\xi_k}-1}{\sigma_k},
\ee
where the case of $\sigma_k=0$ is understood in the limit l'Hopitalle sense
(i.e., for example, $\lambda_k=\xi_k$ if $\sigma_k=0$).

It is easy to see that Eq. (\ref{7}) is satisfied if
\be\label{22}
\nabla^2\xi_k=0,
\ee
i.e. if the functions $\xi_k$ are harmonic. Using the relations (\ref{16})
and Eqs. (\ref{19}), (\ref{20}) one can prove that the Einstein equation
(\ref{8}) reduces to the following one
\be\label{23}
R_{3\,\,\mu\nu}=\frac{1}{4}\sum_k\sigma_k^2\xi_{k,\mu}\xi_{k,\nu}.
\ee
Eqs. (\ref{22}) and (\ref{23}) complete our ansatze definition; finally it is
defined by the relations (\ref{15}), (\ref{16}), (\ref{21})--(\ref{23}). In
this section we will not give a concretezation of the columns $C_k$, functions
$\xi_k$ and metric $h_{\mu\nu}$ and will only mean that the relations
mentioned above are satisfied.

Now let us define the set of vector functions $\vec\nu_k$ on shell of Eq.
(\ref{22}) as
\be\label{24}
\nabla\times\vec\nu_k=\nabla\xi_k.
\ee
Than for the matrix $\vec\Omega$ one obtains the following expression:
\be\label{25}
\vec\Omega=\sum_k\vec\nu_kC_kC_k^T{\cal G}_0,
\ee
see Eqs. (\ref{12}), (\ref{19}), (\ref{20}) and (\ref{24}). To
calculate ${\rm det}\,{\cal G}$ it is convenient to rewrite the matrix
${\cal G}$ in the form of
\be\label{25'}
{\cal G}=\hat{\cal S}{\cal G}_0,
\ee
where the ``evolutionary operator" $\hat{\cal S}$ can be represented as
\be\label{26}
\hat{\cal S}=exp\left ( \sum_k\xi_kC_kC_k^T{\cal G}_0\right ).
\ee
Finally, after the additional application of Eq. (\ref{16}), one obtains
that
\be\label{27}
{\rm det}{\cal G}=exp \left ( \sum_k\sigma_k\xi_k\right ).
\ee
Eqs. (\ref{15}), (\ref{17}), (\ref{21}), (\ref{25}) and (\ref{27}) give the
complete information necessary for calculation of the nonzero components of
the bosonic string theory truncated in section 2. To write down these
components in the explicit form, let us parametrize the columns $C_k$ as
\be\label{28}
C_k=\left (\ba{c}
p_k\cr
q_k
\ea\right ),
\ee
where $p_k$ are the numbers, whereas $q_k$ are the $d\times 1$ columns. Then,
using some algebra and applying Eqs. (\ref{10}), (\ref{11}) and (\ref{13}),
one finally obtains the following result:
\be\label{29}
ds^2_{d+3}&=&dy^T\left ( {\cal G}_0+\sum_kq_kq_k^T\frac{e^{\sigma_k\xi_k}-1}
{\sigma_k}\right )dy+\left ( 1-\sum_kp_k^2\frac{e^{-\sigma_k\xi_k}-1}
{\sigma_k}\right )ds_3^2,\nonumber\\
e^{\Phi}&=&exp\left ( \sum_k\sigma_k\xi_k/2\right ) +\sum_k
p_k^2\frac{\sinh{\sigma_k\xi_k/2}}{\sigma_k/2},\nonumber\\
B_{m,d+\mu}&=&\sum_kp_kq_{km}\nu_{k\mu}.
\ee
Here the constants $p_k$ and $q_k$ must satisfy the relations
\be\label{30}
-p_kp_l+q_k^TG_0q_l=\sigma_k\delta_{kl},
\ee
where $G_0$ is the ``22" block component of the matrix ${\cal G}_0$, i.e.,
$G_0={\rm diag} (-1, 1,...,1)$.

Now let us put ${\cal N}=2$ and restrict the following consideration by
three special cases with $\sigma_1=\sigma_2=\sigma$ and $\sigma=0, \pm 1$.
Let us combine the functional pair $(\xi_1, \xi_2)$ into the single complex
potential
\be\label{31}
\xi=\xi_1+i\xi_2.
\ee
Then the motion equations (\ref{22}) and (\ref{23}) take the following form:
\be\label{32}
\nabla^2\xi=0,
\ee
\be\label{33}
R_{3\,\,\mu\nu}=\frac{\sigma^2}{4}\left ( \xi_{\mu}\bar\xi_{\nu}+
\xi_{\nu}\bar\xi_{\mu}\right ).
\ee
Eqs. (\ref{32}) and (\ref{33}) describe the harmonic complex scalar field
coupled to the 3--dimensional gravity. From Eq. (\ref{33}) one concludes
that $|\sigma|$ plays the role of the effective coupling constant. In the
case of $\sigma=0$ the coupling vanishes and one obtains the flat 3--space
(the extremal case, see also \cite{mpla1}), whereas for $\sigma_k=\pm 1$
this 3--space is curved (the nonextremal case). In the next section we
construct a concrete special solution for the system (\ref{32})--(\ref{33})
and calculate the corresponding complex vector field
\be\label{34}
\vec\nu=\vec\nu_1+i\vec\nu_2,
\ee
which satisfies the relation
\be\label{35}
\nabla\times\vec\nu=\nabla\xi,
\ee
as it follows from Eqs. (\ref{24}) and (\ref{31}). We also give the
explicit expressions for the columns $C_1$ and $C_2$; this last step defines
the multidimensional bosonic string theory fields completely, see Eqs.
(\ref{29}). Thus, in the rest part of this paper we consider the
2--functional subspase of the previously truncated bosonic string theory
written in the compact complex form of Eqs. (\ref{32}), (\ref{33}) and
(\ref{35}).


\section{A solution}

Let $x^{\mu}=(\rho, z, \varphi)$ be the Weil canonical coordinates. We plan
to construct an axisymmetric solution of the system (\ref{32})--(\ref{33}).
In the axisymmetric case $\partial_{\varphi}\sim 0$ on all the functions and
the 3--dimensional metric can be taken in the
Lewis--Papapetrou form \cite{exsol}
\be\label{36}
ds_3^2=e^{\gamma}(d\rho^2+dz^2)+\rho^2d\varphi^2.
\ee
For this metric Eq. (\ref{32}) transforms into the one
\be\label{37}
(\rho\xi_{,\rho})_{,\rho}+(\rho\xi_{,z})_{,z}=0,
\ee
whereas Eq. (\ref{33}) converts into the system
\be\label{38}
\gamma_{,\rho}&=&\frac{\rho}{4}(\xi_{,\rho}\bar\xi_{,z}+
\xi_{,z}\bar\xi_{,\rho}),\nonumber\\
\gamma_{,z}&=&\frac{\rho}{4}(\xi_{,\rho}\bar\xi_{,\rho}-
\xi_{,z}\bar\xi_{,z}).
\ee
From Eq. (\ref{37}) it follows that the 3--space becomes flat in the case
of $\gamma=0$. Then, Eq. (\ref{38}) is $\gamma$--independent, so one can
take an arbitrary axisymmetric solution of the ``usual" flat Laplace equation
as a solution of Eq. (\ref{37}). Then, the system (\ref{38}) is consistent
for the arbitrary solution of Eq. (\ref{37}), so one can calculate the
function $\gamma$ for the arbitrary axisymmetric harmonic $\xi$. We take
\be\label{39}
\xi=\frac{e}{[\rho^2+(z-ia)^2]^{1/2}},
\ee
where
\be\label{40}
e=e_1+ie_2
\ee
and the parameters $e_1, e_2$ and $a$ are the arbitrary real constants.
The harmonic function (\ref{40}) formally corresponds to the complex
Coulomb charge $e$ ``located" on the symmetry axis at the imaginary
``position" $z=ia$. Solving Eq. (\ref{38}), one obtains that
\be\label{41}
\gamma=\frac{|\sigma e|^2}{16a^2}\left \{
1-\frac{\rho^2+z^2+a^2}{[(\rho^2+z^2+a^2)^2-4a^2\rho^2]^{1/2}}
\right \},
\ee
where the integration constant had been chosen to obtain
$\gamma=0$ at the spatial infinity.

To rewrite the solution (\ref{39}), (\ref{41}) in the root--free form, let
us introduce the prolonged spheroidal coordinates
$(r, \theta, \varphi)=x^{\mu}$ accordingly the relations
\be\label{42}
\rho&=&\sqrt{r^2+a^2}\sin\theta,\nonumber\\
z&=&r\cos\theta
.
\ee
Then the function $\xi$ and the metric $ds_3^2$ take the form of
\be\label{43}
\xi=\frac{e}{r-ia\cos\theta},
\ee
\be\label{44}
ds_3^2&=&{\rm exp}\left (
-\frac{|\sigma e|^2}{8}\frac{\sin^2\theta}{r^2+a^2\cos^2\theta}
\right )
\left [
\frac{r^2+a^2\cos^2\theta}{r^2+a^2}dr^2+(r^2+a^2\cos^2\theta)d\theta^2
\right ]+
\nonumber\\
&+&(r^2+a^2)\sin^2\theta d\varphi^2.
\ee
The last step is to calculate the vector field $\vec\nu$ using Eqs.
(\ref{35}), (\ref{43}) and (\ref{44}). The result for the nonzero components
reads:
\be\label{45}
\nu_3=\nu_{\varphi}=e\frac{r\cos\theta-ia}{r-ia\cos\theta}.
\ee
Using Eqs. (\ref{43}) and (\ref{45}) it is easy to obtain the components
$\xi_k$ and $\nu_k$, the result is:
\be\label{46}
\xi_1=\frac{e_1r-e_2a\cos\theta}{r^2+a^2\cos^2\theta},\qquad
\xi_2=\frac{e_1a\cos\theta+e_2r}{r^2+a^2\cos^2\theta},\nonumber
\ee
\be
\nu_{\varphi\,\,1}&=&e_1\cos\theta+a\sin^2\theta
\frac{e_1a\cos\theta+e_2r}{r^2+a^2\cos^2\theta},\nonumber\\
\nu_{\varphi\,\,2}&=&e_2\cos\theta-a\sin^2\theta
\frac{e_1r-e_2a\cos\theta}{r^2+a^2\cos^2\theta}.
\ee
To complete the definition of the bosonic string theory fields, one must
give an explicit expression for the columns $C_1$ and $C_2$. These columns
must yield the quadratic algebraical restrictions (\ref{16}). Solving them,
one obtains that
\be\label{47}
C_1=
\left (\ba{c}
\sqrt{s_1^2-\sigma}\cos(\beta+\frac{\alpha}{2})\cr
\sqrt{s_1^2-\sigma}\sin(\beta+\frac{\alpha}{2})\cr
s_1n_1
\ea\right ),
\quad
C_2=
\left (\ba{c}
\sqrt{s_2^2-\sigma}\cos(\beta-\frac{\alpha}{2})\cr
\sqrt{s_2^2-\sigma}\sin(\beta-\frac{\alpha}{2})\cr
s_2n_2
\ea\right ),
\ee
where
\be\label{48}
n_1^Tn_1=n_2^Tn_2=1
\ee
and
\be\label{49}
\cos\alpha=\frac{s_1s_2}{\sqrt{(s_1^2-\sigma)(s_2^2-\sigma)}}n_1^Tn_2,
\ee
whereas the angle parameter $\beta$ is arbitrary.

Eqs. (\ref{29}), (\ref{44}) and (\ref{46})--(\ref{49}) completely define the
constructed solution. Let us now discuss some its properties. First of all,
from Eqs. (\ref{29}) and (\ref{46}) it follows that the multidimensional
metric is asymptotically flat, i.e., it describes the Minkowskian space--time
at $r\rightarrow\infty$. Then, the dilaton field vanishes at the spatial
infinity, so there is a sense to define and compute its charge. We define the
dilaton charge by the relation
\be\label{50}
\Phi\sim\frac{D}{r}, \quad r\rightarrow\infty;
\ee
then from Eqs. (\ref{29}) and (\ref{46}) it follows that
\be\label{51}
D=e_1(p_1^2+\frac{\sigma}{2})+e_2(p_2^2+\frac{\sigma}{2}).
\ee
Taking into account Eq. (\ref{47}), one finally concludes that
\be\label{52}
D=e_1[(s_1^2-\sigma)\cos^2(\beta+\frac{\alpha}{2})+\frac{\sigma}{2}]+
e_2[(s_2^2-\sigma)\cos^2(\beta-\frac{\alpha}{2})+\frac{\sigma}{2}].
\ee
Then, from Eqs. (\ref{29}) and (\ref{46}) it follows that the constructed
solution possesses the Dirac string peculiarity for the Kalb--Ramond field
components $B_{m,d+3}$. The corresponding term does not vanish at
$r\rightarrow\infty$ and contains the NUT--like (\cite{nut})
$r$--independent component
\be\label{53}
B_{m,d+3}\sim (e_1p_1q_{1m}+e_2p_2q_{2m})\cos\theta.
\ee
To remove this peculiarity and to obtain the completely asymptotically
flat solution one must restrict the solution parameters by the relation
\be\label{54}
e_1p_1q_{1}+e_2p_2q_{2}=0.
\ee
After some algebra one concludes that Eq. (\ref{54}) can not be satisfied
in the case of $\sigma=1$ if the Kalb--Ramond field remains nontrivial.
For the case of $\sigma=0$ the solution of Eq. (\ref{54}) reads:
\be\label{55}
s_1^{-1}C_1= s_2^{-1}C_2=
\left (\ba{c}
\cos\beta\cr
\sin\beta\cr
n
\ea\right ),
\ee
where $n^Tn=1$. In this case
\be\label{55'}
e_1=\frac{s_2}{s_1}e_0, \quad e_2=-\frac{s_1}{s_2}e_0,
\ee
where the parameter $e_0$ is arbitrary; from Eq. (\ref{51}) it follows that
$D=0$ here. For the case of $\sigma=-1$ one obtains three distinct special
situations:

a) \, the ``symmetric" charge configuration with $e_1=e_2=e_0$, $D=0$ and
\be\label{56}
C_1=
\left (\ba{c}
\cos\beta\cr
\sqrt{1+s^2}\sin\beta\cr
sn\sin\beta
\ea\right ),
\quad
C_2=
\left (\ba{c}
-\sin\beta\cr
\sqrt{1+s^2}\cos\beta\cr
sn\cos\beta
\ea\right );
\ee

b) \, the ``nonsymmetric" charge configuration with $e_1=e_0$, $e_2=0$,
$D=-e_0/2$ and
\be\label{57}
C_1=\cosh\beta_1
\left (\ba{c}
0\cr
1\cr
n\,\tanh\beta_1
\ea\right ),
\quad
C_2=\cosh\beta_2
\left (\ba{c}
\frac{\sqrt{1+\sinh^2\beta_1+\sinh^2\beta_2}}{\cosh\beta_1\cosh\beta_2}\cr
\tanh\beta_1\tanh\beta_2\cr
n\,\tanh\beta_2
\ea\right );
\ee

c) \, and the another "nonsymmetric" charge configuration which can be
obtained from the previous one using the replacement $1\leftrightarrow 2$.

Finally, Eqs. (\ref{44}), (\ref{46}) together with Eqs. (\ref{47})--(\ref{49})
(in the case of the possible Dirac string existence) and Eqs.
(\ref{55})--(\ref{57}) (in the Dirac peculiarity free case) give the total
concretezation of the solution expressed by Eq. (\ref{29}). We leave the
study of its regularity as well as the horizon analysis to the forthcoming
articles.


 \section*{Acknowledgments}
 This work was supported by RFBR grant ${\rm N^{0}}
 \,\, 00\,02\,17135$ and by CONACYT grant I32799-E.


 \section{Conclusion}

In this paper we have constructed a special class of asymptotically flat
solutions of the consistently truncated effective field theory of the bosonic
string. This class belongs to the two--functional solution subspace which
admits a compact complex form. Our concrete solution is related to the
special complex Coulomb--like axisymmetric solution of the 3--dimensional
Laplace equation. Our physical interpretation of the obtained solution is
based on its asymptotical behaviour near to the spatial infinity. It is shown
that using an appropriate choice of the solution parameters one can remove
a Dirac string and obtain a solution describing the point--like source with
the nontrivial Kalb--Ramond dipole moment and the dilaton charge vanishing in
the extremal case and possessing both zero and nonzero values for the
nonextremal configurations. The constructed solution class can be transformed
into the corresponding one for the nonstatic Kaluza--Klein theory coupled to
the dilaton field using the duality relations given in \cite{mpla1}. Moreover,
in the case of $d=2$ it is possible to rewrite our solutions as the solutions
of the complete 5--dimensional bosonic string theory compactified on a
3--torus. In the framework of the $(d+3)$--dimensional effective string
theories one can apply the charging symmetry subgroup of U--dualities
\cite{cs}
to generate both the bosonic and heterotic string theory degrees of freedom
``missed" in our present ``truncated" consideration.

Let us now discuss the performed analysis and its possible generalizations.
In fact, the crucial steps had been done in section 3, where we have
introduced the matrices
\be\label{c1}
\Pi_{kk}=C_kC_k^T.
\ee
These matrices form a set of the mutually orthogonal projectional
operators; this statement is supported by the list of their products.
In the considered case of $k=1,2$ this list reads:
\be\label{c2}
\Pi_{11}^2=\sigma_1\Pi_{11}, \quad \Pi_{22}^2=\sigma_2\Pi_{22}, \quad
\Pi_{11}\Pi_{22}=0.
\ee
In this paper we have taken the equal signs $\sigma_1=\sigma_2=\sigma$, and
have studied the dependence of the Dirac string removing in our concrete
solution on this common sign.
In the context of the scheme presented it will be interesting to develop the
general formalism based on the use of ${\cal N}$ arbitrary columns $C_k$ and
to relate the solution subspace with the corresponding projection operators
$\Pi_{kl}=C_kC^T_l$. In this general situation one works again with the
closed table of the mutual operator products, and obtains a real possibility
to calculate the nontrivial asymptotically flat solutions of the type more
general than presented in this paper. It will be interesting to establish
the general role of this projectional formalism in the framework of the
theories possessing the $\sigma$--model representation.

As the nearest interesting perspective it will be important to develop the
projectional formalism in the situation, where the general toroidally
compactified heterotic string theory will be projected to the stationary and
axisymmetric Einstein--Maxwell theory. This problem seems solvable in the
above discussed generalized projectional approach and, in the case of its
realization, opens new wide possibilities in extension of the all known
solutions of the Einstein--Maxwell theory, including supersymmetric ones
\cite{t}, to the (super)string theory field.


 \end{document}